\documentclass[fleqn,twoside]{article}
\usepackage{espcrc2}

\usepackage{graphicx}
\usepackage[figuresright]{rotating}

\newcommand{\AmS}{{\protect\the\textfont2
  A\kern-.1667em\lower.5ex\hbox{M}\kern-.125emS}}

\title{Optical Clocks in Space}

\author{
S. Schiller\address{Institut f{\"u}r Experimentalphysik, Universit\"{a}t D\"{u}sseldorf, Universit\"atsstr.\,1, 40225 D\"{u}sseldorf, Germany}\thanks{Proc. III International Conference on Particle and Fundamental Physics in Space (SpacePart06), Beijing 19 - 21 April 2006, to appear in Nucl. Phys. B}, 
A. G\"orlitz\addressmark, A. Nevsky\addressmark, J.C.J. Koelemeij\addressmark, A. Wicht\addressmark,
P. Gill\address{National Physical Laboratory, Teddington, Middlesex TW11 0LW,
United Kingdom}, H. A. Klein\addressmark, H. S. Margolis\addressmark,
G. Mileti\address{Observatoire de Neuch{\^a}tel, 58, rue de l'Observatoire
CH-2000 Neuch{\^a}tel, Switzerland},
U. Sterr\address{Physikalisch-Technische Bundesanstalt, Bundesallee 100,
38116 Braunschweig, Germany}, F. Riehle\addressmark, E. Peik\addressmark, Chr. Tamm\addressmark,
W. Ertmer\address{Institut f\"ur Quantenoptik, Leibniz  Universit\"at Hannover,
Welfengarten 1, 30167 Hannover, Germany}, E. Rasel\addressmark, 
V. Klein\address{Kayser-Threde GmbH, Wolfratshauser Str. 48, 81379 M\"u nchen,
Germany},
C. Salomon\address{
Ecole Normale Sup{\'e}rieure - D\'ep. de Physique,
24 rue Lhomond, 75231 Paris Cedex 05, France}, 
G.M. Tino\address{Dip. di Fisica / Lab. LENS,
  Univ. di Firenze,
via Sansone 1, 50019 Sesto Fiorentino,
  Italy},
P. Lemonde\address{LNE -SYRTE, Observatoire de Paris, 61 av. de l'Observatoire, 75014 Paris,
France},
R. Holzwarth\address{Max-Planck-Institut f\"ur Quantenoptik, Garching, Germany}, T.W. H\"ansch\addressmark
}

\begin{document}

\begin{abstract}
The performance of optical clocks has strongly progressed in recent years, and 
accuracies and instabilities of 1 part in 10$^{18}$ are expected in the near future. The operation of optical clocks in space
provides new scientific and technological opportunities. In particular, an earth-orbiting satellite containing
an ensemble of optical clocks would allow a precision measurement of the gravitational redshift, navigation with improved precision,
mapping of the earth's gravitational potential by relativistic geodesy, and comparisons between ground clocks.
\vspace{1pc}
\end{abstract}

\maketitle

\section{Introduction}

Earth-based operation of future optical clocks \cite{EFTF06}
 will be limited by uncertainties in the value of the gravitational potential $U$ at the clocks' locations and in their relative velocity. For example, an uncertainty $\delta U/U=1\cdot10^{-9}$ (equivalent to a 1\,cm elevation change) and a relative velocity of 1\,cm/year, caused, e.g. by continental drift, would lead to a gravitational redshift and a first-order Doppler shift contribution in the frequency comparison amounting to 1 part in $10^{18}$. For applications requiring the highest time and frequency precision it is therefore essential to operate optical clocks in space, where the above effects can be determined with sufficient accuracy, by continuously measuring the orbit parameters.

The following applications of optical clocks in space are of interest:

- distribution of time and frequency on earth and in space from an earth-orbiting "master clock"

- mapping of the earth's gravity field by frequency comparison of terrestrial clocks with the 
master clock. The terrestrial clocks are transported over land or sea to cover areas. This method will complement terrestrial clock-clock comparisons using optical fibers.

- precision spacecraft navigation using the master clock signals

- space-based VLBI

- metrological applications, e.g. space gravitational-wave interferometers

- high precision tests of gravitational physics

\section{Tests of the gravitational redshift}

From the perspective of fundamental physics, the most exciting possibility is to test the Einstein prediction of the gravitational redshift with extremely high accuracy \cite{Schiller05}. 
The gravitational redshift between two spatially separated clocks of equal frequency $\nu$ is given by $(\nu_1-\nu_2)/\nu=(U({\bf r}_2)-U({\bf r}_1))/c^2$, to lowest order in $c^{-1}$. Higher-order contributions in $c^{-1}$ must be taken into account 
\cite{Linet02}.

 A satellite-based measurement of the absolute gravitational redshift could be done in two ways. First, a comparison between clocks on satellites in different circular orbits, including possibly a geostationary orbit. Then the gravitational potential difference and thus the gravitational redshift would be approximately constant in time, and therefore one needs to rely on achieving a high absolute accuracy of the clocks in space. Alternatively, as proposed for the mission concept OPTIS \cite{Lämmerzahl}, one satellite is on a highly elliptical orbit, while the second clock ($\nu_{ter}$) is located on earth. The latter's gravitational shift is not relevant since it is sufficiently constant on the orbit timescale.  Since the frequency shift $\nu_{sat}-\nu_{ter}$ is modulated in time at the orbital frequency, the requirement on the clocks is a high stability on the orbital period timescale.  This second approach is simpler. A high accuracy of the clocks is still of interest for the other applications listed above. In both cases,
$U(t)$ for the satellite clock(s) must be accurately determined, e.g. by  laser ranging of the satellite(s).

On an eccentric earth orbit with apogee and perigee on the order of 36.000\,km and 10.000\,km, respectively, the gravitational potential variation is $\Delta U/c^2\simeq 2\cdot10^{-10}$. With clock instabilities and time transfer at the level of $1\cdot10^{-18}$ on the orbital half-period ($\simeq 7\,$h) a test of the redshift with relative accuracy of 1 part in $10^8$ per orbit, and better than 1 part in 10$^9$ after averaging over a year would be possible. 
   
A second test is a test of the universality of the gravitational redshift, a consequence of the principle of local position invariance (LPI). LPI asserts that the outcome of nongravitational local experiments is independent of where the experiment is performed. Thus, a comparison of the frequencies of two clocks of different nature located in the same satellite should yield the same value, no matter where the satellite is with respect to massive bodies. An LPI test is simpler than an absolute gravitational redshift test because it is performed entirely within the satellite. An accurate knowledge of $U$ along the orbit is not necessary. Again, the crucial clock requirement is low instability, not high accuracy.

The outcome of any experiment can in principle be expressed in terms of the dimensionless fundamental constants and other experiment-specific dimensionless parameters such as number of particles involved. Assuming the latter strictly constant, we may describe violations of LPI as arising from a dependence of some fundamental constant $\beta$ on the gravitational potential, $D_\beta\equiv  d(ln\beta)/d(U/c^2)\ne0$. These coefficients should be tested for as many fundamental constants as possible.  Using (conventional) optical clocks, one can test the $U$-dependence of the fine structure constant $\alpha$ and of the electron-to-nucleon mass  ratio $m_e/m_N$, via the frequencies of atomic electronic transitions and of vibrational molecular transitions, respectively. A comparison of an electronic ($\nu_{at}$) with a vibrational ($\nu_{vib}$) clock at the same location and experiencing a change $\Delta U$ (e.g. by clock transport) may be expressed as
\begin{displaymath}
{\Delta({\nu_{at}/\nu_{vib}})\over {\nu_{at}/\nu_{vib}}}=(b_\alpha D_\alpha+b_{\hat m_e} D_{\hat m_e}+
\end{displaymath} 
\hskip 1in $b_{\hat m_q} D_{\hat m_q}+b_{\hat m_s} D_{\hat m_s}{)}{\Delta U\over c^2}\ .$

\vskip.05in
\noindent Here, ${\hat m_e}=m_e/\Lambda_{QCD}$, ${\hat m_q}=m_q/\Lambda_{QCD}$, ${\hat m_s}=m_s/\Lambda_{QCD}$ are the masses of electron, light and strange quark relative to the QCD energy scale, respectively, and the coefficients $b$ depend on the particular transitions chosen. Choosing suitable transitions leads to $b_\alpha,\,b_{\hat m_e} \simeq1$. The values
 $b_{\hat m_q}=-0.037\, b_{\hat m_e}$, $b_{\hat m_s}= -0.011\, b_{\hat m_e}$ are related because they arise from $D_{m_e/m_N}$ 
\cite{Flambaum}. A comparison of two electronic transitions (in the same or in different atoms) with differing $\alpha$-dependence can be used for testing for the single parameter $D_\alpha$. From vibrational transitions it is not possible to individually determine $D_{\hat m_e}$, $D_{\hat m_q}$, and $D_{\hat m_s}$. This requires using additional clock types, such as hyperfine transition clocks, whose dependence on the nuclear $g$-factor introduces different dependencies on the quark masses.

An LPI test with relative accuracy $1\cdot10^{-9}$ will be possible with the assumed optical clocks. This is almost 5 and 6 orders more accurate than the best past tests of $D_\alpha$ and $D_{m_e/m_N}$, respectively. At this accuracy,  the universality at a level of second order in $\Delta U/c^2$ may become relevant.

Since the optical clocks will require clock lasers stabilized to optical cavities, these can be used for performing a test of the isotropy of light propagation, an aspect of the principle of Local Lorentz Invariance. The near-zero gravity environment would be advantageous since distortions of the cavities would be very small. The satellite would have to spin at a highly constant rate to allow the test. 
 
\section{Implementation issues}

A suitable satellite payload could consist of an ion clock, a cold neutral atom clock, and a molecular clock.
It is advantageous that space-qualified optical subsystems similar to those needed for such a clock ensemble have already been developed, for example laser-pumped vapour cell clocks, the cold-atom clock PHARAO, laser communication and lidar systems. In detail, these systems include components such as: single-frequency diode lasers, ultracold atom sources, opto-electronic components, solid-state lasers and amplifiers, optical resonators, phase-locked lasers.

Femtosecond frequency combs and optical clocks for future space use are currently under study.


Accurate time transfer between the satellite clock and the receiver, be it on earth or on another satellite, is critical: a level better than 1 part in $10^{18}$ over several hours is required in order to take advantage of clock performance. New time transfer technology beyond the advanced microwave link for ACES  will have to be developed. Possibly, the shift from microwaves to optical waves could be necessary. A demonstration of time transfer from earth to the low-earth orbit satellite JASON via laser link (T2L2) is planned in 2008 \cite{Samain05}. It will be based on pulsed lasers with $\simeq 10\,$ns pulse duration, but its peformance will still be far from the above requirement.  The use of fs laser pulses and coherent optical links may represent an alternative approach for enhanced accurary. The feasibility of continuous-wave coherent optical links for information transfer has already been demonstrated in long-distance atmospheric experiments \cite{Horwath06} and for space application a study is planned in 2007 (LCT experiment on TerraSAR).  A transportable optical ground station has been developed. Such stations would be of importance for comparing  transportable optical clocks to space clocks. 

\noindent{\it Acknowledgments} European agencies are supporting research towards the development of optical frequency metrology for space, in particular through studies MAP AO 2004-100, 19595/05/NL/PM,
AO/1-5057/06/F/VS. 
We gratefully acknowledge their support.

\end{document}